\documentclass[twocolumn,aps,pra,showpacs,groupedaddress,superscriptaddress,nofootinbib,raggedbottom]{revtex4-1}
\usepackage{graphicx}  
\usepackage{color}     
\usepackage{inputenc}  
\usepackage{amsmath}   
\usepackage{amsfonts}  
\usepackage{float}     
\usepackage[pdftex,linkcolor=blue,citecolor=blue,urlcolor=blue,colorlinks]{hyperref} 

\begin{document}

\title{Orbital angular momentum dynamics of Bose-Einstein condensates \\trapped in two stacked rings}
\author{Eul\`{a}lia Nicolau}
\author{Jordi Mompart}%
\affiliation{%
	Departament de F\'{i}sica, Universitat Aut\`{o}noma de Barcelona, E-08193 Bellaterra, Spain
}%

\author{Bruno Juli\'{a}-D\'{i}az}%
\affiliation{Departament de F\'{i}sica Qu\`{a}ntica i Astrof\'{i}sica, Facultat de F\'{i}sica, Universitat de Barcelona, 08028 Barcelona, Spain.}%
\affiliation{Institut de Ci\`{e}ncies del Cosmos, Universitat de Barcelona, ICCUB, Mart\'{i} i Franqu\`{e}s 1, Barcelona 08028, Spain}%

\author{Ver\`{o}nica Ahufinger}%
\affiliation{%
	Departament de F\'{i}sica, Universitat Aut\`{o}noma de Barcelona, E-08193 Bellaterra, Spain
}%

\begin{abstract}

We investigate the stability and dynamics of the orbital angular momentum modes of a repulsive Bose-Einstein condensate trapped in two tunnel-coupled rings in a stack configuration. Within mean-field theory, we derive a two-state model for the system in the case in which we populate equally both rings with a single orbital angular momentum mode and include small perturbations in other modes. Analyzing the fixed point solutions of the model and the associated classical Hamiltonian, we characterize the destabilization of the stationary states and the subsequent dynamics. By populating a single orbital angular momentum mode with an arbitrary population imbalance between the rings, we derive analytically the boundary between the regimes of Josephson oscillations and macroscopic quantum self-trapping and study numerically the stability of these solutions.  

\end{abstract}
\maketitle

\section{Introduction}
Ultracold atoms trapped in ring potentials are one of the most promising systems in the emerging field of atomtronics \cite{Seaman2007,Amico2017}. They have been considered for quantum sensing applications such as rotation sensing \cite{Sherlock2011,Arnold2012}, magnetometry \cite{Pelegri2018}, Sagnac interferometry \cite{Barrett2014,Helm2015,Moxley2016,Nolan2016,Navez2016,Safaei2019}, or the atomic analogue to the superconducting quantum interference devices (SQUIDs) \cite{Anderson2003,Ramanathan2011,Wright2013,Ryu2013,Amico2014,Eckel2014,Eckel2014b,Jendrzejewski2014,Wang2015}. Rings are the simplest geometries that lead to non-trivial loop circuits, in which the superfluidity of Bose-Einstein condensates (BECs) gives rise to persistent currents \cite{Beattie2013,Corman2014}. One can transfer orbital angular momentum (OAM) to the trapped BEC either by rotating a weak link \cite{Ramanathan2011} or by coherent transfer of angular momentum from photons to the atoms \cite{Andersen2006}. Regarding the implementation of the ring trapping potential, several techniques have been implemented or proposed: magnetic traps \cite{Gupta2005}, optically plugged magnetic traps \cite{Ryu2007}, conical refraction \cite{Turpin2015}, pairs of optical fibers \cite{Moscatelli2007}, static Laguerre-Gauss Beams \cite{Wright2000}, and time-averaged  \cite{Sherlock2011,Arnold2012,Bell2016} or painting \cite{Schnelle2008,Henderson2009} potentials. 

On the other hand, the Josephson effect is a fundamental phenomenon in quantum mechanics that has been widely explored in superconductors, and its study has been recently extended to bosonic ultracold atomic systems \cite{Josephson1962,Anderson1963,Smerzi1997,Albiez2005,Shin2005,Levy2007}. Josephson oscillations can arise in weakly coupled BECs trapped in a double-well potential: when there is a non-zero population imbalance, quantum tunneling allows the particles to oscillate periodically from one well to the other. However, repulsive interactions can suppress tunneling such that the atoms remain mostly trapped in one of the wells, regime known as macroscopic quantum self trapping \cite{Smerzi1997}. Weakly coupled condensates have been proposed as basic building blocks for quantum technologies \cite{Schumm2005,Hall2007,Jo2007,Esteve2008}. In particular, the dynamics of BECs in tunnel-coupled ring potentials have been thoroughly explored in a variety of geometries such as stacked rings with \cite{Aghamalyan2013,Escriva2019} or without lattices \cite{Lesanovsky2007,Brand2010,Oliinyk2019,Oliinyk2019b,Oliinyk2020}, concentric rings \cite{Zhang2012,Polo2016}, or coplanar rings \cite{Polo2016a,Bland2019}.

In this work, we investigate a BEC trapped in two rings in a stack configuration to study the interplay between the OAM, the tunneling dynamics, and the repulsive nonlinear interactions. First, we consider an initial state with a single OAM mode equally populated in both rings, which gives rise to symmetric and antisymmetric stationary states. The stability conditions for these states against OAM perturbations were derived within the mean-field theory and using Bogoliubov analysis in \cite{Brand2010}. Here, we revisit the problem and demonstrate that the system can be described by a two-state model with fixed point solutions. In particular, one can derive a classical Hamiltonian that characterizes the dynamics of the system in terms of the orbits around the critical points. Second, we consider an initial state where a single OAM mode is populated with a non-zero population imbalance between rings, such that tunneling and interactions give rise to different dynamical regimes. We derive analytically the boundary condition between Josephson oscillations and self-trapping, and study numerically the stability of these regimes against perturbations in higher order OAM modes. 

The paper is organized as follows. In Section \ref{SecPhysicalSystem}, we describe the physical system and introduce the few-state model of OAM modes derived from the Gross-Pitaevskii equation. Section \ref{SecStability} deals with the stability of the stationary states: after presenting briefly the Bogoliubov analysis, we derive a two-state model, find its critical points and analyze its associated classical Hamiltonian. The model is then compared against numerical simulations of the complete system of equations derived in Section \ref{SecPhysicalSystem}. Section \ref{SecDynamics} focuses on the dynamical regimes of Josephson oscillations and self-trapping: we first study the case of populating a single mode in each ring and then explore the role of higher order OAM perturbations. Finally, the conclusions are presented in Section \ref{SecConclusions}.
\section{Physical system}\label{SecPhysicalSystem}
\begin{figure}[t]
	\center
	\includegraphics[width=0.31\textwidth]{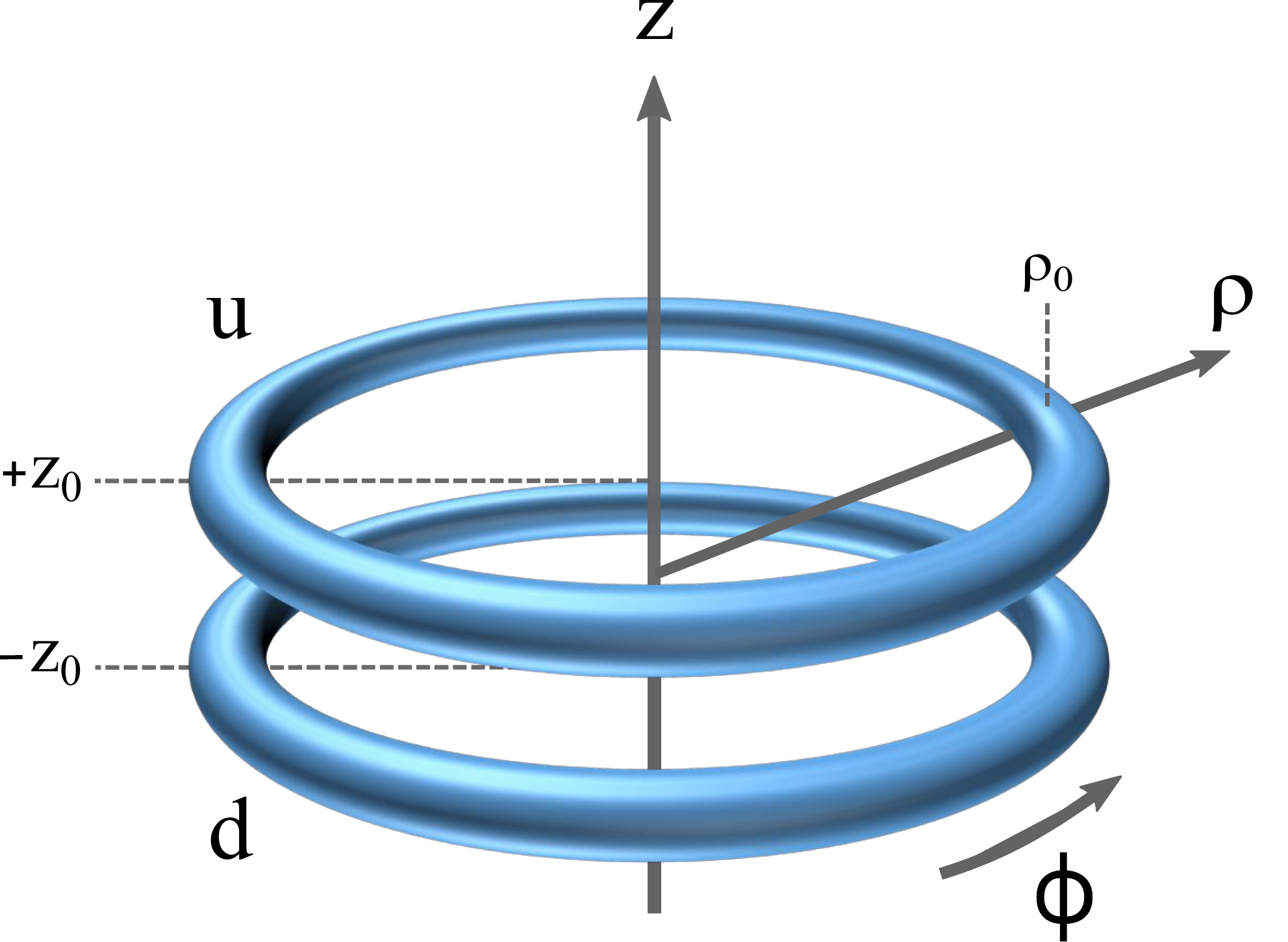}
	\caption{Schematic representation of the geometry of the system. The trapping potential consists of two ring traps, up, $u$, and down, $d$, that are located in the planes $\pm z_0$, centered at $\rho=0$, and have radius $\rho_0$.}\label{FigSistema}
\end{figure}

The system under consideration is shown in Fig.~\ref{FigSistema}. It consists of two coaxial annular traps around the $z$-axis separated by a distance $2z_0$, where a BEC of $N$ atoms is trapped. The BEC is described within the mean-field theory by the Gross-Pitaevskii equation (GPE), which in cylindrical coordinates reads
\begin{eqnarray}
i\hbar\dfrac{\partial \Psi(\textbf{r},t) }{\partial t}&=&\bigg[\dfrac{\hbar^2}{2M}\Big(-\dfrac{\partial^2}{\partial\rho^2}-\dfrac{1}{\rho}\dfrac{\partial}{\partial\rho}-\dfrac{\partial^2}{\partial z^2}+ \dfrac{L_z^2}{\hbar^2\rho^2}\bigg)+\nonumber\\ 
&&\quad+V(\textbf{r})+g|\Psi(\textbf{r},t)|^2\Big]\Psi(\textbf{r},t), \label{grosscylindrical}
\end{eqnarray}
where $V(\textbf{r})$ is the external potential, $M$ is the atomic mass, $L_z=-i\hbar\frac{\partial}{\partial\phi}$ is the $z$ component of the angular momentum, and $g=4\pi\hbar^2a_s/M$ accounts for the contact interactions characterized by the $s$-wave scattering length $a_s$. The wavefunction, $\Psi(\textbf{r},t)$, is normalized to the total number of particles, $N$. Henceforth, we will consider exclusively repulsive interactions, $g>0$, and rings with large enough radii such that the term $\frac{1}{\rho}\frac{\partial}{\partial\rho}$ can be neglected in Eq.~(\ref{grosscylindrical}). The trapping potential in (\ref{grosscylindrical}) is defined as $V(\textbf{r})=V_z(z)+V_\rho(\rho)$, where $V_z$ is a symmetric double-well harmonic potential with minima at $\pm z_0$, and $V_\rho$ is a harmonic radial potential centered at $\rho_0$. We assume weak coupling between the rings and that $V_z$ and $V_\rho$ are steep enough such that the BEC only presents azimuthal excitations. Then, the wavefunction can be factorized as
\begin{equation}\label{completewavefunction}
\Psi(\textbf{r},t)=\Psi(\rho)\left[\Phi^u(z)\chi^u(\phi,t)+\Phi^d(z)\chi^d(\phi,t)\right],
\end{equation}
where $\Psi(\rho)$ is the ground state of the radial harmonic potential and the functions $\chi^{u}(\phi,t)$ and  $\chi^{d}(\phi,t)$ contain the dependence of the BEC wavefunction with respect to time. The functions $\Phi^{u}(z)$ and $\Phi^{d}(z)$ are two modes localized in the wells up ($u$) and down ($d$) constructed as a superposition of the ground and first excited stationary solutions of the GPE equation. The total number of particles in each ring is $\int d\phi|\chi^{u/d}(\phi,t)|^2=N^{u/d}(t)$ and the functions $\Psi(\rho)$, $\Phi^u(z)$ and $\Phi^d(z)$ are normalized to $1$. 
The functions $\chi^{u}(\phi,t)$ and $\chi^{d}(\phi,t)$ for the upper and lower rings can be written as a linear combination of the angular momentum eigenstates, 
\begin{equation}\label{chiProportional}
\chi^{u/d}(\phi,t)=\frac{1}{\sqrt{2\pi}}\sum_{m=-\infty}^{\infty} \alpha_{m}^{u/d}(t)\hspace{1mm}e^{im\phi},
\end{equation}
with amplitudes $\alpha_{m}^{u/d}(t)$. For each eigenstate, the condensate has a quantized angular momentum $m\hbar$. The angular mode coefficients are normalized to the number of particles in the $m$-th angular mode in each ring, $|\alpha_{m}^{u/d}(t)|^2=N_{m}^{u/d}(t)$, such that $N^{u/d}(t)=\sum_m N_m^{u/d}(t)$. Henceforth, we will omit the explicit time dependence in $\alpha_{m}^{u/d}(t)$. The evolution equations for the amplitudes of each OAM mode, $\alpha_m^{u/d}$, read \cite{Lesanovsky2007,Brand2010}: 
\begin{equation}\label{sistemEquations}
i\dfrac{\partial \alpha_m^{u/d}}{\partial \tau}\hspace{-0.5mm}=\hspace{-0.5mm}m^2\alpha_m^{u/d}-\kappa\alpha_m^{d/u}+\gamma\sum_{nn'}\alpha_n^{u/d}(\alpha_{n'}^{u/d})^*\alpha_{m-n+n'}^{u/d},
\end{equation}
where $\tau=\hbar t/(2MR^2)$ is the scaled time, $\kappa=R^2\int\hspace{-0.9mm} dz\, (\Phi^d(z))^*\big[\frac{\partial^2}{\partial z^2}-\frac{2M}{\hbar^2}V_z\big]\Phi^u(z)$ is the tunneling rate between the two rings, and $\gamma=MR^2g/(\pi\hbar^2)\int d\rho \rho|\Psi(\rho)|^4 \int dz |\Phi^u(z)|^4$  is the interatomic interaction parameter with $R^{-2}=\int d\rho\rho^{-1}|\Psi(\rho)|^2$. The first term of the RHS in (\ref{sistemEquations}) corresponds to the kinetic energy of the $m$-th mode, the second term, to the tunneling between the two rings, which only couples OAM modes with the same $m$, and the third term is the nonlinear interaction that couples different OAM modes within each ring. 


\section{Stability of the stationary states}\label{SecStability}
Let us consider that only one OAM mode $n$ is initially populated in both rings: $|\alpha_{n}^{u/d}(\tau=0)|^2\neq0$, $|\alpha_{m\neq n}^{u/d}(\tau=0)|^2=0$. Then, stationary solutions only exist for equal number of particles between rings, $N_{n}^{u}=N_{n}^{d}= N/2$, and Eq.~(\ref{sistemEquations}) simplifies to
\begin{equation}
i\dot{\alpha}_{n}^{u/d}=n^2\alpha_{n}^{u/d}-\kappa\alpha_{n}^{d/u}+\epsilon\alpha_{n}^{u/d},
\end{equation}
where $\epsilon=\gamma N/2$ and the dot indicates the derivative with respect to $\tau$. By diagonalizing this system of equations, we find the following symmetric and antisymmetric stationary solutions with energies $\mu_\pm$:
\begin{subequations}\label{EqStationaryStates}
	\begin{align}
	\label{EqSymmetricSolution}
	\hspace{-1mm}(\alpha_n^u,\alpha_n^d)_s&=\sqrt{N}e^{-i\mu_+\tau}(1,1), \, & \mu_+&=n^2+\epsilon-\kappa\\
	\label{EqAntisymetricSolution}\hspace{-1mm}(\alpha_n^u,\alpha_n^d)_a&=\sqrt{N}e^{-i\mu_-\tau}(1,-1), \, & \mu_-&=n^2+\epsilon+\kappa.
	\end{align}
\end{subequations}

\subsection{Bogoliubov analysis}\label{SecBogoliubov}
\begin{figure}[!tp]
	\includegraphics[width=0.47\textwidth]{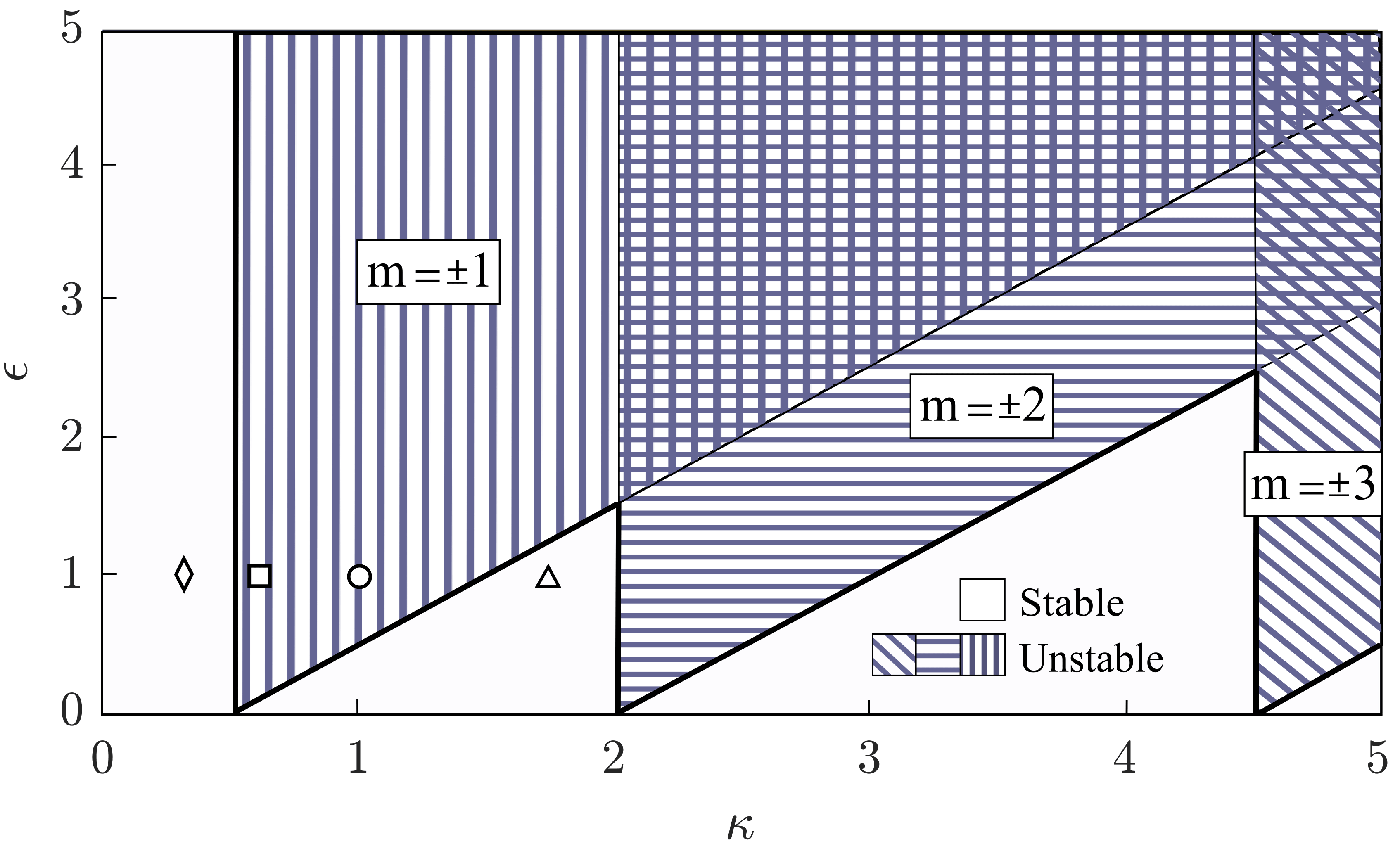}
	\caption{Real (white) and imaginary (patterned) regions of $\omega$ for the antisymmetric state with $n=0$ and perturbations in the modes $m=\pm1,\pm2,\pm3$ in the parameter space $[\kappa,\epsilon]$. The points rhombus, square, circle and triangle correspond to the parameter values used in Fig.~\ref{FigPathsMap} with the circle also being used in Fig.~\ref{comparison}.}\label{S}
\end{figure}

In order to study the stability of the states (\ref{EqSymmetricSolution}) and (\ref{EqAntisymetricSolution}), we fix $n=0$ and add a small amplitude symmetric perturbation in an arbitrary mode $m\neq0$, of the form
\begin{equation}\label{ansatzBogoliubov}
\alpha_{ m}^{u/d}=e^{-i\mu_\pm\tau}(u_{m}^{u/d}e^{-i\omega\tau}+(v_{m}^{u/d})^*e^{i\omega\tau}).
\end{equation}
By introducing this ansatz together with (\ref{EqStationaryStates}) into (\ref{sistemEquations}) and linearizing for small amplitudes of $u_{m}^{u/d}$ and $(v_{m}^{u/d})^*$, we obtain the following Bogoliubov-de Gennes equations
\begin{subequations}\label{bogoEigenvalueEq}
\begin{align}
\label{EqBogoU}
&\omega u_{m}^{u/d}=(m^2-\mu_\pm+2\epsilon)u_{m}^{u/d}+\epsilon v_{-m}^{u/d}-\kappa u_{m}^{d/u}\\
\label{EqBogoV}
-&\omega v_{-m}^{u/d}=(m^2-\mu_\pm+2\epsilon)v_{-m}^{u/d}+\epsilon u_{m}^{u/d}-\kappa v_{-m}^{d/u}.
\end{align}
\end{subequations}
By diagonalizing (\ref{bogoEigenvalueEq}), one finds that only the antisymmetric state can be unstable against perturbations in higher order modes. The corresponding excitation branch, $\omega$, determines the regions of the parameter space for which the antisymmetric state is unstable \cite{Brand2010}:
\begin{equation}\label{excitationbranch}
\omega=\sqrt{(m^2+\epsilon-2\kappa)^2-\epsilon^2}.
\end{equation}
For real values of $\omega$, the perturbations (\ref{ansatzBogoliubov}) remain periodic and thus bounded, while, for imaginary values, the perturbations in mode $m$ grow exponentially, destabilizing the stationary state. Fig.~\ref{S} shows the real (white) and imaginary (patterned) regions of $\omega$ for the stationary state with $n=0$ and perturbations in $m=\pm1,\pm2, \pm3$ as a function of $\kappa$ and $\epsilon$. Interactions increase the instability regions of the antisymmetric state. The spectrum in (\ref{excitationbranch}) also holds for stationary solutions with $n\neq0$; in that case, the perturbation $m$ is the OAM difference with respect to $n$. 

\subsection{Two-state model}\label{SecTwoStateModel}
The Bogoliubov analysis predicts the stability regions of the stationary solutions in the parameter space. However, it does not describe the dynamics once the stationary state has been destabilized. In order to get an insight into the excitation process, we derive the simplest model that captures these dynamics: a two-state model that includes the antisymmetric stationary state mode and a pair of perturbation modes $\pm m$. We take for simplicity the mode $n=0$ for the stationary state, with $|\alpha_{0}^{u}|^2=N_{0}^{u}$ and $|\alpha_{0}^{d}|^2=N_{0}^{d}$. Then, the system of equations (\ref{sistemEquations}) reduces to a set of six equations that can be expressed in matrix form as 
\begin{equation}\label{EqMatrixSistem}
i\begin{pmatrix}
\dot{\alpha}_{0}^u\\
\dot{\alpha}_{m}^u\\
\dot{\alpha}_{-m}^u\\
\dot{\alpha}_{0}^d\\
\dot{\alpha}_{m}^d\\
\dot{\alpha}_{-m}^d\\
\end{pmatrix} =\hat{A}\cdot
\begin{pmatrix}
\alpha_{0}^u\\
\alpha_{m}^u\\
\alpha_{-m}^u\\
\alpha_{0}^d\\
\alpha_{m}^d\\
\alpha_{-m}^d\\
\end{pmatrix},\end{equation}
where the matrix $\hat{A}$ reads
\begin{widetext}
\begin{equation}\label{totalmatrix}
\mbox{\small$\displaystyle
	\begin{aligned}
	\begin{pmatrix}
	\gamma(2N^{u}{-}|\alpha_{0}^{u}|^2) & \gamma\alpha_{-m}^u(\alpha_0^u)^* & \gamma\alpha_{m}^u(\alpha_0^u)^* & {-}\kappa & 0 & 0 \\
	\gamma\alpha_{0}^u(\alpha_{-m}^u)^* & \hspace{-1mm}m^2{+}\gamma(2N^{u}{-}|\alpha_{m}^{u}|^2) & 0 & 0 &{-}\kappa & 0\\
	\gamma\alpha_{0}^u(\alpha_{m}^u)^*  & 0  & \hspace{-1mm}m^2{+}\gamma(2N^{u}{-}|\alpha_{-m}^{u}|^2) & 0 & 0 & {-}\kappa \\
	{-}\kappa & 0 & 0 & \hspace{-1mm}\gamma(2N^{d}{-}|\alpha_{0}^{d}|^2) & \gamma\alpha_{-m}^d(\alpha_0^d)^* & \gamma\alpha_{m}^d(\alpha_0^d)^* \\
	0 & {-}\kappa & 0 & \gamma\alpha_{0}^d(\alpha_{-m}^d)^* & \hspace{-1mm}m^2{+}\gamma(2N^{d}{-}|\alpha_{m}^{d}|^2) & 0 \\
	0 & 0 & {-}\kappa & \gamma\alpha_{0}^d(\alpha_{m}^d)^* & 0 & \hspace{-1mm}m^2{+}\gamma(2N^{d}{-}|\alpha_{-m}^{d}|^2)\\
	\end{pmatrix},
		\end{aligned}$}
\end{equation}
with $N^{u/d}=|\alpha_0^{u/d}|^2+|\alpha_{m}^{u/d}|^2+|\alpha_{-m}^{u/d}|^2$ being the total number of particles in the $u$ and $d$ rings. We impose the initial condition $\alpha_{0}^u=-\alpha_0^d$ and add small amplitude symmetric perturbations in the high order modes $\pm m$ such that $\delta\alpha_{\pm m}^{u}=\delta\alpha_{\pm m}^{d}$. 
Due to angular momentum conservation and the fact that the stationary state is in the mode $n=0$, the conditions $|\alpha_{m}^{u}|^2=|\alpha_{-m}^{u}|^2$ and $|\alpha_{m}^{d}|^2=|\alpha_{-m}^{d}|^2$ are fulfilled. 
Assuming that the phase difference between the perturbed modes stays approximately constant during the time evolution and that $|\alpha_{\pm m}^d|\approx|\alpha_{\pm m}^u|$, we can define $\alpha_{ m}\equiv\alpha_{\pm m}^u=\alpha_{\pm m}^d$. We will also assume that the initial condition $\alpha_{0}^u=-\alpha_0^d$ is maintained during the temporal evolution, so that we can also use $N^u\approx N^d=N/2$. Applying all these conditions, the expression~(\ref{EqMatrixSistem}) can be simplified to a set of three equations for $\alpha_0^u$, $\alpha_{m}$ and $\alpha_0^d$, that in matrix form  reads: 
\begin{equation}
i\hspace{-0.7mm}\begin{pmatrix}
\dot{\alpha}_{0}^u\vspace{1.6mm}\\
\dot{\alpha}_{m}\vspace{1.6mm}\\
\dot{\alpha}_{0}^d\\
\end{pmatrix}\hspace{-1mm}=\hspace{-1mm}
\begin{pmatrix}
\gamma\big(N-|\alpha_{0}^{u}|^2\big(1-2\big(\frac{\alpha_{m}}{\alpha_{0}^{u}}\big)^2\big)\big) &0 & -\kappa \\
0 & \hspace{-1mm}-\kappa+m^2+\gamma\big(N-|\alpha_{ m}^{}|^2\big(1-\big(\frac{\alpha_{0}^{u}}{\alpha_{m}^{}}\big)^2\big)\big)  & 0 \\
-\kappa & 0  & \hspace{-1mm}\gamma\big(N-|\alpha_{0}^{u}|^2\big(1-2\big(\frac{\alpha_{ m}^{}}{\alpha_{0}^{u}}\big)^2\big)\big) 
\end{pmatrix}
\hspace{-1.5mm}\begin{pmatrix}
\alpha_{0}^u\vspace{1.6mm}\\
\alpha_{m}\vspace{1.6mm}\\
\alpha_{0}^d\\
\end{pmatrix}.
\end{equation}
This system can be reduced further by noting that the first and last diagonal elements are equal. Then, defining $\alpha_0\equiv\alpha_{0}^u$ we obtain the following two-state model:
\begin{equation}\label{EqTwoStateModelMatrix}
\begin{aligned}
i\begin{pmatrix}
\dot{\alpha}_0\vspace{1.3mm}\\
\dot{\alpha}_{m}
\end{pmatrix}=\begin{pmatrix}
\gamma(N-|\alpha_{0}|^2\big(1-2\big(\frac{\alpha_{m}}{\alpha_{0}}\big)^2\big)\big)+\kappa &0  \\
0 & -\kappa+m^2+\gamma\big(N-|\alpha_{m}^{}|^2\big(1-\big(\frac{\alpha_{0}}{\alpha_{m}^{}}\big)^2\big)\big)  
\end{pmatrix}\begin{pmatrix}
\alpha_0\vspace{1.3mm}\\
\alpha_{m}
\end{pmatrix}.
\end{aligned}
\end{equation}
\end{widetext}

In order to understand the oscillatory dynamics of the system, we define  $\alpha_0=|\alpha_0|e^{i\phi}$ and $\alpha_m=|\alpha_m|e^{i\theta}$. By using particle conservation,  $2|\alpha_0|^2+4|\alpha_m|^2=N$, and defining the phase difference $\zeta=\theta-\phi$, the system reduces to two coupled real equations: 
\begin{subequations}
	\begin{align}
	\dot{|\alpha_m|^2}&=2\gamma |\alpha_m|^2 \left(2|\alpha_m|^2-\dfrac{N}{2}\right)\sin2\zeta\label{Eq2StateModel1}\\
	\begin{split}
	\dot{\zeta}&=2\kappa-m^2+\gamma\left(3|\alpha_m|^2-\dfrac{N}{2}\right)\label{Eq2StateModel2}+\\
	&\hspace{7mm} +\gamma\left(4|\alpha_m|^2-\dfrac{N}{2}\right)\cos{2\zeta}.
	\end{split}
	\end{align}
\end{subequations}

\subsubsection{Critical points}
The critical points of this system fulfill $\dot{|\alpha_m|^2}=\dot{\zeta}=0$. Imposing $\dot{|\alpha_m|^2}=0$ in Eq.~(\ref{Eq2StateModel1}), we find
\begin{equation}\label{Eq.StationaryPoints}
|\alpha_m|^2=0, \qquad |\alpha_m|^2=\dfrac{N}{4}, \qquad \sin 2\zeta=0,
\end{equation}
where the two first trivial solutions correspond to the minimum and maximum values of $|\alpha_m|^2$ that are due to particle conservation. The critical points can be then found imposing $\dot{\zeta}=0$ in Eq.~(\ref{Eq2StateModel2}). For the trivial cases, the critical points are 
\begin{subequations}
\begin{align}
\bigg( \cos{2\zeta}&=\dfrac{2\kappa-m^2-\epsilon}{\epsilon}\equiv\cal{A},&|\alpha_m|^2&=0\bigg) \label{EqCritical point0} \\ 
\bigg(\cos{2\zeta}&=\dfrac{m^2-2\kappa-\epsilon/2}{\epsilon}\equiv\cal{B},&|\alpha_m|^2&=\dfrac{N}{4}\bigg).\label{EqCritical pointN/4} 
\end{align}
\end{subequations}
Due to the boundedness of the cosine in (\ref{EqCritical point0}), the solution with $|\alpha_m|^2=0$ exists if
\begin{equation}\label{EqCondition1}
\dfrac{m^2}{2}\leq \kappa\leq \dfrac{m^2+2\epsilon}{2},
\end{equation}
and similarly, the solution with $|\alpha_m|^2=N/4$, Eq.~(\ref{EqCritical pointN/4}), exists if
\begin{equation}\label{EqCondition2}
\dfrac{m^2-3\epsilon/2}{2} \leq \kappa\leq \dfrac{m^2+\epsilon/2}{2}.
\end{equation}
By studying the eigenvalues of the Jacobian at the critical points, these trivial solutions can be shown to be saddle points (see Appendix).\\

For the nontrivial solution, for which $|\alpha_m|^2$ takes values different from $0$ or $N/4$, the critical points are 
\begin{subequations}\label{EqCenterPosition}
\begin{align}
\bigg(\zeta&=a\pi,&|\alpha_m|^2=&\dfrac{m^2-2\kappa+2\epsilon}{14\epsilon/N}\equiv\cal{C}\bigg) \label{EqCritical Point pi}\\ 
\bigg(\zeta&=(2a+1)\dfrac{\pi}{2},&|\alpha_m|^2=&\dfrac{2\kappa-m^2}{2\epsilon/N}\equiv\cal{D}\bigg), \label{EqCritical point pi/2}
\end{align}
\end{subequations}
where $a\in \mathbb{Z}$. Taking into account the minimum and maximum values of $|\alpha_m|^2$ due to particle conservation, the solutions with $\zeta=a\pi$ exist if
\begin{equation}\label{EqCondition3}
\dfrac{m^2-3\epsilon/2}{2} \leq \kappa\leq \dfrac{m^2+2\epsilon}{2},
\end{equation}
whereas the ones with $\zeta=(2a+1)\pi/2$ exist if 
\begin{equation}\label{EqConditionRestrictive}
\dfrac{m^2}{2} \leq \kappa\leq \dfrac{m^2+\epsilon/2}{2}.
\end{equation}
Note that the second set of solutions, Eq.~(\ref{EqCritical point pi/2}), has a more restrictive condition than the first, Eq.~(\ref{EqCritical Point pi}). Similarly as before, these solutions can be shown to be centers, with the trajectories orbiting around them (see Appendix).

\subsubsection{Two-state model Hamiltonian}
\begin{figure}[!htb]
	\center
	\includegraphics[width=0.49\textwidth]{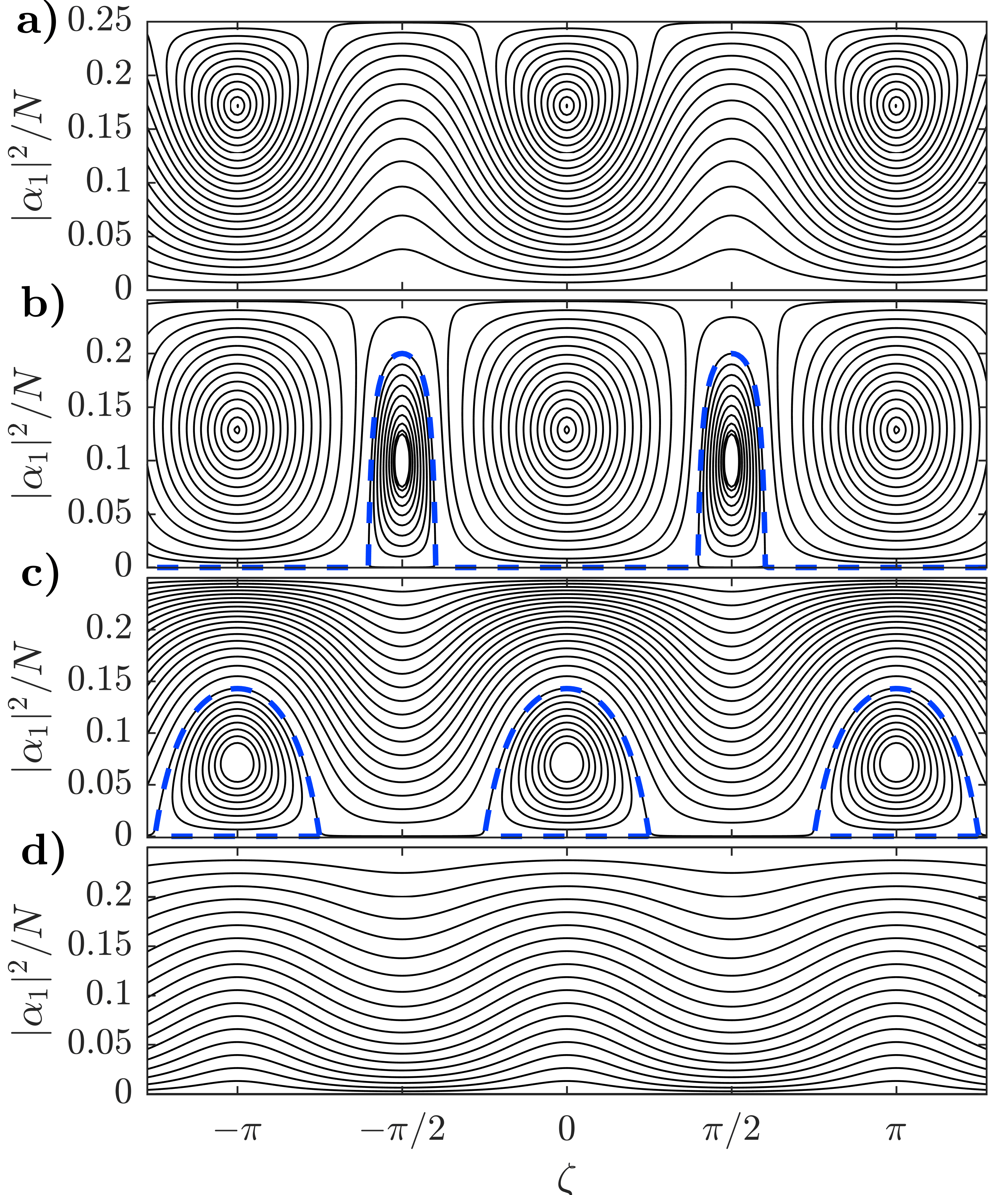}
	\caption{Lines of constant $H(|\alpha_1|^2,\zeta)$ for $m=1$, $\gamma=1/2000$, and $N=4000$ (thus, $\epsilon=1$) and a) $\kappa=0.3$, b) $\kappa=0.6$, c) $\kappa=1$, d) $\kappa=1.7$. In dashed blue, orbits corresponding to the stationary state excitations for the unstable cases, b) and c). The parameter values of the plots correspond to the points rhombus, square, circle and triangle in Fig.~\ref{S}, respectively.}\label{FigPathsMap}
\end{figure}

Assuming that the variables $|\alpha_m|^2$ and $\zeta$ are canonical conjugates, they fulfill $\partial H/\partial (|\alpha_m|^2)=\dot{\zeta}$ and $\partial H/\partial \zeta=-\dot{|\alpha_m|^2}$, and thus the corresponding classical Hamiltonian $H$ reads:
\begin{equation}\label{Eq2StateHamiltonian}
\begin{aligned}
H(|\alpha_m|^2,\zeta)=
& |\alpha_m|^2\Bigg[2\kappa-m^2-\dfrac{\gamma N}{2}+\dfrac{3}{2}\gamma |\alpha_m|^2+\\
&+\gamma\left(2|\alpha_m|^2-\dfrac{N}{2}\right)\cos{2\zeta}\Bigg].
\end{aligned}
\end{equation}
Fig.~\ref{FigPathsMap} shows lines of constant $H(|\alpha_1|^2,\zeta)$ for various initial conditions and $\gamma=1/2000$, $N=4000$ (thus, $\epsilon=1$), $m=1$ and different values of $\kappa$. According to the existence conditions of the critical points, Eqs.~(\ref{EqCondition1},\ref{EqCondition2},\ref{EqCondition3},\ref{EqConditionRestrictive}), there are four possible types of phase diagrams as a function of the tunneling $\kappa$:
\begin{itemize}
	\item $(m^2-3\epsilon/2)/2<\kappa<m^2/2$: there are saddle points at ($\cal{B}$, $|\alpha_m|^2=N/4$) and centers at ($\zeta=a\pi$, $\cal{C}$) (\textit{e.g.} Fig.~\ref{FigPathsMap}(a)). The orbits around the centers are not accessible for the initial conditions $|\alpha_m|^2/N\simeq 0$ and $\zeta=0$, thus, the stationary state is stable. 
	\item $m^2/2<\kappa<(m^2+\epsilon/2)/2$: there are saddle points at ($\cal{A}$, $|\alpha_m|^2=0$) and ($\cal{B}$, $|\alpha_m|^2=N/4$), and centers at ($\zeta=a\pi$, $\cal{C}$) and ($\zeta=(2a+1)\pi/2$, $\cal{D}$) (\textit{e.g.} Fig.~\ref{FigPathsMap}(b)). Given Eq.~(\ref{EqCenterPosition}), the value of $|\alpha_m|^2$ corresponding to the centers at $\zeta=a\pi$ diminishes with the tunneling $\kappa$ while the one for the centers at $\zeta=(2a+1)\pi/2$ grows with $\kappa$. For the values of $\kappa$ when the $|\alpha_m|^2$ value of the centers at $\zeta=a\pi$ is equal or inferior than those of $\zeta=(2a+1)\pi/2$, the system orbits around ($\zeta=a\pi$, $\cal{C}$). For lower values of $\kappa$, the contrary occurs, and the system performs open orbits around the centers ($\zeta=(2a+1)\pi/2$, $\cal{D}$) (\textit{e.g.} Fig.~\ref{FigPathsMap}(b)).
	\item  $(m^2+\epsilon/2)/2<\kappa<(m^2+2\epsilon)/2$: there are saddle points at ($\cal{A}$, $|\alpha_m|^2=0$) and centers at ($\zeta=a\pi$, $\cal{C}$) (\textit{e.g.} Fig.~\ref{FigPathsMap}(c)), which allows the system to perform orbits around these centers.
	\item For all other values of $\kappa$, \textit{i.e.}, $(m^2+2\epsilon)/2<\kappa<(m^2-3\epsilon/2)/2$, there are neither saddle points nor centers (\textit{e.g.} Fig.~\ref{FigPathsMap}(d)), such that the stationary state is stable.
\end{itemize} 
Combining all these conditions we find that the antisymmetric stationary state is unstable for $m^2/2\leq\kappa\leq(m^2+2\epsilon)/2$, which coincides with the stability conditions predicted by the Bogoliubov analysis (see Fig.~\ref{S}). The Bogoliubov excitations correspond to the open and closed orbits around the centers given the initial conditions $\zeta=0$ and $|\alpha_m|^2/N\simeq 0$, as the ones shown in  blue dashed lines in Figs.~\ref{FigPathsMap}(b) and \ref{FigPathsMap}(c).

The population transfer between the states with $n=0$ and the perturbations $m$ during the excitation is determined by the corresponding orbit.  One can find an upper bound to the population transfer, $|\alpha_m|^2_{\rm max}/N$, by considering the initial conditions $\zeta(\tau=0)=0$ and $|\alpha_m(\tau=0)|^2/N=0$, which correspond to the orbit with $H(|\alpha_m|^2,\zeta)=0$. Taking into account the different possible orbits, either open or closed, and particle conservation in Eq.~(\ref{Eq2StateHamiltonian}), one reaches
\begin{equation}\label{EqPopulationTransfer}
\dfrac{|\alpha_m|^2_{\rm max}}{N}\hspace{-0.5mm}=\hspace{-0.5mm}\left\{
\begin{array}{ll}\hspace{-1mm}\dfrac{2\kappa-m^2}{\epsilon}; & \dfrac{m^2}{2}\leq \kappa\leq\dfrac{m^2+\epsilon/4}{2}\vspace{1.5mm}\\
\hspace{-1mm}\dfrac{2}{7}\dfrac{m^2-2\kappa+2\epsilon}{2\epsilon}; & \dfrac{m^2+\epsilon/4}{2}\leq\kappa\leq \dfrac{m^2+2\epsilon}{2}.
\end{array}\right.
\end{equation}
The upper bound of the population transfer grows linearly with the tunneling $\kappa$, and reaches its maximum for $\kappa=\frac{\epsilon/4+m^2}{2}$, when the centers at $\zeta=(2a+1)\pi/2$ and $\zeta=a\pi$ have the same $|\alpha_m|^2$. Then, the upper bound of the population transfer decreases linearly with $\kappa$ down to zero.


For an initial state with $n\neq0$, one observes analogous dynamics as the ones described above, where the pairs of excited modes have an OAM difference $\pm m$ with respect to $n$. For example, for $\kappa=1$, $\epsilon=1$ and the stationary state with $n=0$, the states that form the excitation are $m=\pm1$, whereas for $n=1$, the excited modes are $m=0$ and $m=2$.

\subsection{Numerical simulations}\label{SecNumericalExample}
\begin{figure}[!tb]
	\centering
	\includegraphics[width=0.48\textwidth]{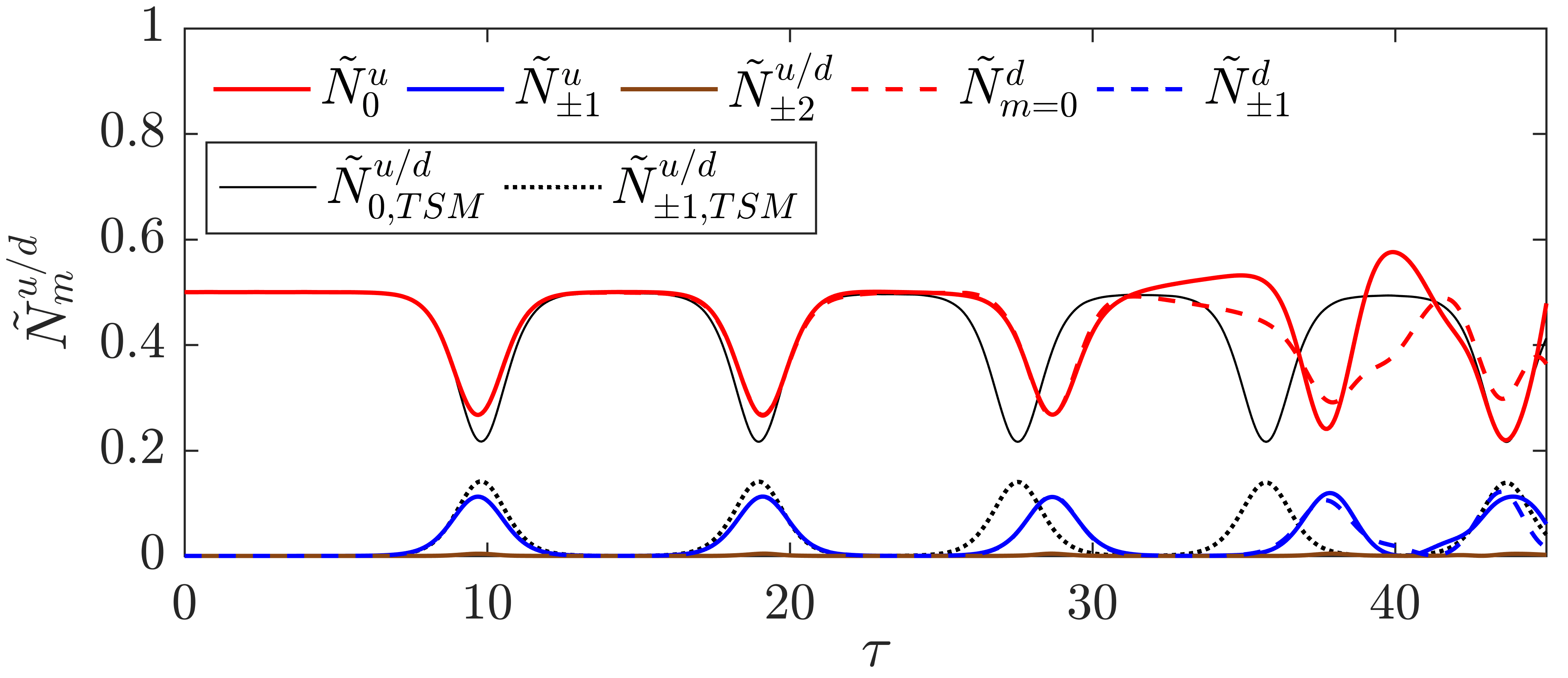}
	\caption{(Color online). Temporal evolution of the populations, $\tilde{N}_m^{u/d}=N_m^{u/d}/N$, for $N=4000$, $\kappa=1$ and $\epsilon=1$ (circle in Fig.~\ref{S}) of the two-state model with $m=1$ (black) and the complete system of equations up to $m=\pm15$ (color). Initial conditions: $\alpha_{0}^u=-\alpha_{0}^d=\sqrt{N/2}$ with  perturbations of order $\sqrt{N}\cdot10^{-4}$ (up to $m=\pm5$ for the complete system of equations (\ref{sistemEquations})). }\label{comparison}
\end{figure}

In this section, we will compare numerically the predictions of the two-state model~(\ref{EqTwoStateModelMatrix}) and the complete system of equations~(\ref{sistemEquations}) for the stationary state with $n=0$, $\kappa=1$ and $\epsilon=1$ (corresponding to the circle in Fig. ~\ref{S}). Fig.~\ref{comparison} shows the time evolution of the populations according to the two-state model (black) and by numerical integration of the system of equations (color). 

For the two-state model, we initially set the amplitudes to $\alpha_0=\sqrt{N/2}$ and $\alpha_1=\sqrt{N}\cdot10^{-4}$ in the system of equations (\ref{EqTwoStateModelMatrix}). The population of the perturbation $\alpha_1$ grows exponentially, in agreement with Eq.~(\ref{excitationbranch}) of the Bogoliubov analysis. Then, the growth of the perturbation slows down, the population reaches a maximum closely bounded by Eq.~(\ref{EqPopulationTransfer}), and the transfer of population is inverted; the population returns to $\alpha_0$. This population transfer pattern is repeated periodically and, for small $\tau$, it precisely captures the dynamics predicted by the complete set of equations. 

For the full model, we populate equally the $n=0$ modes, $\alpha_{0}^{u}=-\alpha_{0}^{d}=\sqrt{N/2}$, and introduce perturbations of order $\sqrt{N}\cdot10^{-4}$ for $m\neq 0$ up to $m=\pm 5$ in Eq. (\ref{sistemEquations}). We include the first $m=\pm15$ modes in the simulation, thus truncating the system of equations well above the highest relevant mode. In this case, the excitation is formed by the pair of modes $m=\pm1$, which evolve with the same population within each ring, thus conserving angular momentum. For long times, the periodic pattern in the evolution of the populations is no longer accurately described by the two-state model since the system does not keep the same population in the $n=0$ modes of the two rings. However, the variations in the period and amplitude of the oscillations could be explained using the two-state model, which suggests that the dynamics of the system are highly sensitive to perturbations (see Fig. 3) i.e., a small perturbation can cause the system to change the orbit. Thus, by analogy, the perturbations appearing during the evolution in the full model would lead to oscillations presenting small changes in their period and amplitude. Also, the maximum population that the excitations reach is lower than the one of the two-state model due to secondary excitations: the higher order modes that are also excited modify the dynamics of the main excitation,   $m=\pm1$. In this case, the mode $m=\pm2$ reaches populations of order $O(10^{-3})$ while higher order modes have smaller contributions.  

\section{Dynamical regimes}\label{SecDynamics}
Thus far, we have studied the destabilization of the stationary states, which have a single OAM mode $n$ populated with the same number of particles in both rings. However, when the initial population in each ring is not the same, tunneling and interactions give rise to different dynamical regimes in the system. 

\begin{figure}[!tb]
	\center
	\includegraphics[width=0.48\textwidth]{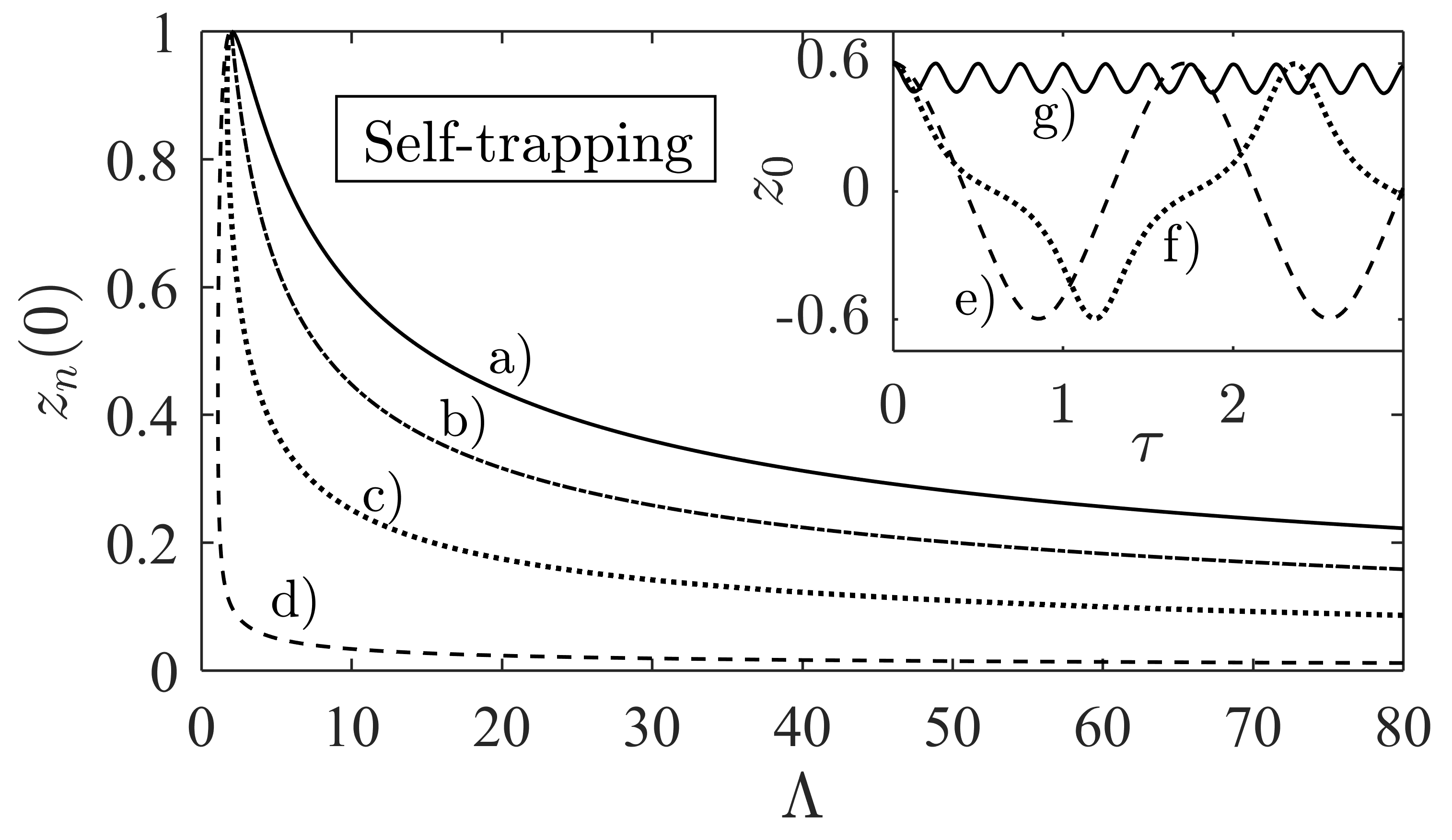}
	\caption{Boundary between the self-trapping and the Josephson oscillations regimes as predicted by (\ref{EqBoundary}) for: a) $\delta\phi_n(0)=0$, b) $\delta\phi_n(0)=\pi/2$, c) $\delta\phi_n(0)=3\pi/4$, d) $\delta\phi_n(0)\rightarrow\pi$. Inset: time evolution of $z_0$ for  $z_0(0)=0.6$, $\delta\phi_0(0)=0$, $N=4000$, $\kappa=1$ and e) $\Lambda=4$, f) $\Lambda=10$, g) $\Lambda=24$.}\label{FigBoundary}
\end{figure}

The dynamics of BECs trapped in double-well potentials are known to present either Josephson oscillations or self-trapping depending on the ratio between the tunneling and the nonlinear interaction \cite{Albiez2005}.  In the Josephson oscillations regime, the population performs complete oscillations between the two wells while in the self-trapping regime, the population remains mostly trapped in one well. In order to find the self-trapping condition for our system, we initially populate a single mode $n$ and factorize the amplitudes as \vspace{0.5mm}$\alpha_{n}^{u/d}=\sqrt{\vphantom{N^{\dot{P}}}{\smash{\small{N_{n}^{u/d}}}}}e^{i\beta_{n}^{u/d}}$. The system of equations~(\ref{sistemEquations}) can then be rewritten in terms of the population imbalance, $z_m=(N_m^u-N_m^d)/N$, and the phase difference, $\delta\phi_{n}=\beta_{n}^d-\beta_{n}^u$, as a set of two coupled equations:
\begin{equation}
\left\{
	\begin{aligned}
 \dot{z}_{n}&=-\sqrt{1-z_{n}^2}\sin{\delta\phi_{n}} \\
 		\delta\dot{\phi}_{n}&=\Lambda z_{n}+ \frac{z_{n}}{\sqrt{1-z_{n}^2}}\cos{\delta\phi_{n}},
	\end{aligned}\right.
\end{equation}
where $\Lambda=\gamma N/(2\kappa)=\epsilon/\kappa$ and $\tau$ has been scaled to $2\kappa\tau$. Assuming that $z_{n}$ and $\delta\phi_{n}$ are canonically conjugate variables, then, $\partial H/\partial z_{n}=\delta\dot{\phi}_{n}$ and $\partial H/\partial \delta\phi_{n}=-\dot{z}_{n}$, and the corresponding classical Hamilonian reads
\begin{equation}\label{EqHamiltonian}
\begin{aligned}
H=\frac{1}{2}\Lambda z_{n}^2-\cos{\delta\phi_{n}}\sqrt{1-z_{n}^2}.
\end{aligned}
\end{equation}
Note that the Hamiltonian is equal for all $n$. Thus, the system presents identical dynamics for all OAM modes. In order to find the boundary between the regimes of self-trapping and Josephson oscillations, we impose $z_{n}(\tau)=0$, which is only fulfilled in the Josephson oscillations regime.  Using energy conservation in (\ref{EqHamiltonian}) and denoting the initial parameters as $z_{n}(\tau=0)\equiv z_n(0)$ and $|\delta\phi_{n}(\tau=0)|\equiv\delta\phi_n(0)$, one reaches
\begin{equation}\label{EqBoundary}
\Lambda_c=2\Bigg(\frac{\cos{\delta\phi_n(0)}\sqrt{1-z_n^2(0)}+1}{z_n^2(0)}\Bigg),
\end{equation}
which defines the phase boundary between the two regimes in terms of the initial population imbalance and phase difference and the nonlinear interactions. This condition is a generalization of the one found in \cite{Smerzi1997} for a BEC in a double-well potential. Fig.~\ref{FigBoundary} shows the boundary given by (\ref{EqBoundary}) for different values of the initial phase difference $\delta\phi_n(0)$ as a function of $\Lambda=\epsilon/\kappa$ and the initial population imbalance $z_{n}(0)$. The self-trapping regime occurs for sufficiently large imbalance and ratio $\Lambda=\epsilon/\kappa$. As the phase difference grows from $0$ to $\pi$, the region of parameters for which self-trapping occurs grows and, as one approaches the limit $\delta\phi_n(0)\rightarrow\pi$, the minimum population imbalance to obtain self-trapping approaches $z_{n}(0)=0$.

The inset in Fig.~\ref{FigBoundary} shows the temporal evolution of the population imbalance, $z_0$, for $z_{0}(0)=0.6$, $\delta\phi_n(0)=0$ and for different values of $\Lambda$: e) $\Lambda=4$, f) $\Lambda=10$, g) $\Lambda=24$. 
As the ratio $\Lambda=\epsilon/\kappa$ grows, the oscillations become anharmonic until the average population imbalance becomes non-zero. If one further increases $\Lambda$, the amplitude of the remaining oscillations decreases and they are eventually suppressed, thus the population remains at the initial imbalance. 

\subsection{Stability of the dynamical regimes}\label{SecDynamicalInstability}
In this section we study numerically the stability of the dynamical regimes, Josephson oscillations and self-trapping, in the presence of perturbations in higher order modes. Initially, we populate the mode $n=0$ with a certain imbalance $z_0(0)$ between the rings and a phase difference of $\pi$, and introduce small amplitude perturbations in higher order modes of order $\sqrt{N}\cdot10^{-4}$ for $m\neq 0$ up to $m=\pm 3$. 

\begin{figure}[!tb]
	\includegraphics[width=1\columnwidth]{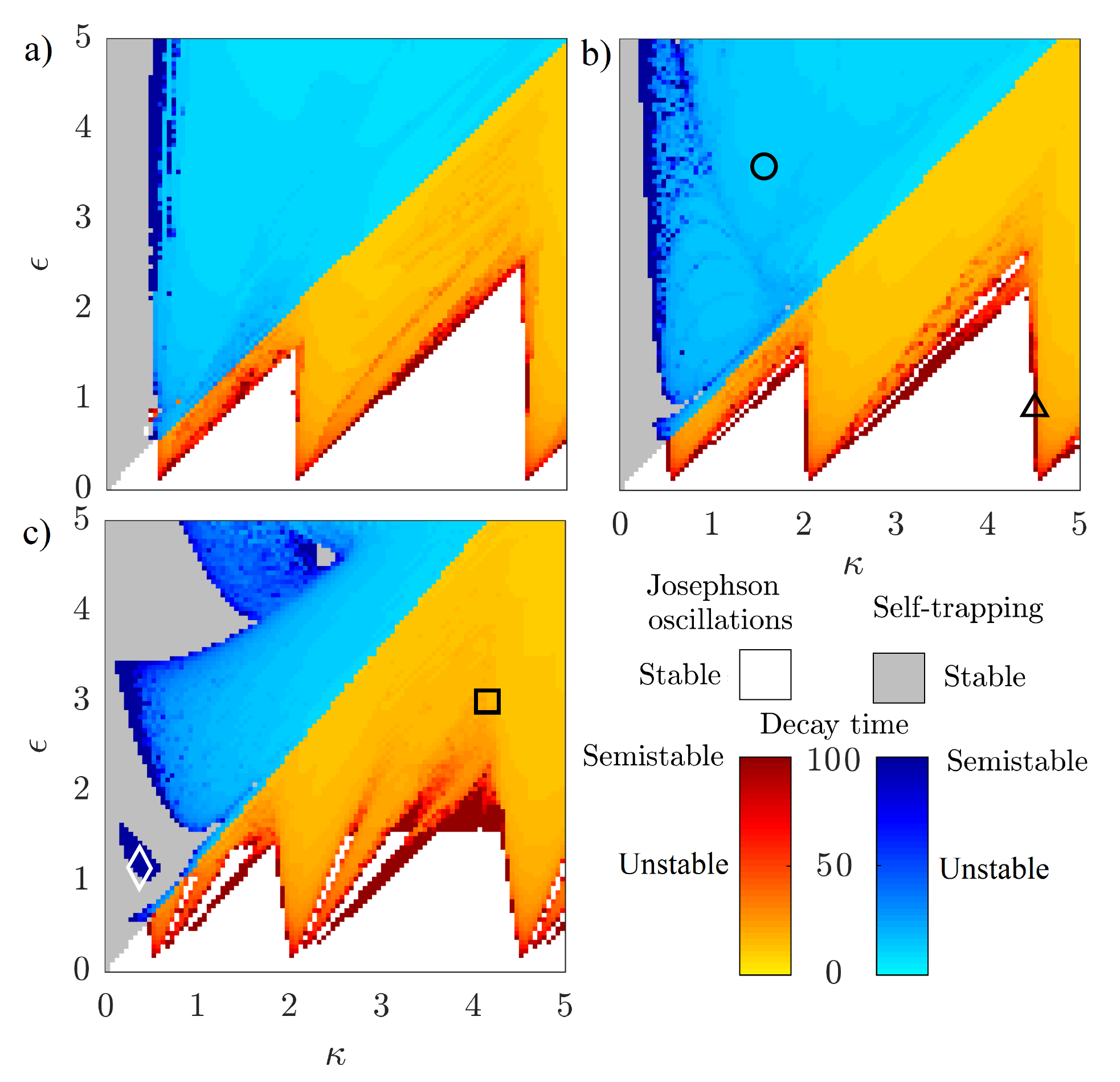}
	\caption{(Color online). Dynamical regimes in the parameter space [$\kappa,\epsilon$] up to $\tau=100$ with $N=4000$ for a) $z_0(0)=0.1$, b) $z_0(0)=0.4$, and c) $z_0(0)=0.75$. The marked points correspond to: square (Fig.~\ref{FigUnstable}(a)), circle (Fig.~\ref{FigUnstable}(b)), triangle (Fig.~\ref{FigJOsemistable}) and white rhombus (Fig.~\ref{FigSTsemistable}). For the semistable cases, the dynamics do not decay up to $\tau=100$.}\label{FigMaps}
\end{figure}

Fig.~\ref{FigMaps} shows the different dynamics in the parameter space [$\kappa,\epsilon$] for a) $z_0(0)=0.1$, b) $z_0(0)=0.4$ and c) $z_0(0)=0.75$. Grey and blue indicate stable and unstable self-trapping, respectively, and white and red indicate stable and unstable Josephson oscillations. The simulations run up to $\tau=100$, and darker shades of blue and red indicate longer decay times. The boundary between Josephson oscillations and self-trapping is not modified by the perturbations, and thus it is determined by Eq.~(\ref{EqBoundary}) taking $\delta\phi_0(0)=\pi$. For small initial imbalance, Eq.~(\ref{excitationbranch}) predicts accurately the regions of stability of the dynamical regimes, as the initial state resembles the stationary state (see Figs.~\ref{S} and \ref{FigMaps}(a)). As the initial imbalance gets larger, the structure of the unstable regions becomes more involved (Figs.~\ref{FigMaps}(b) and \ref{FigMaps}(c)). 

\begin{figure}[!b]
	\includegraphics[width=1\columnwidth]{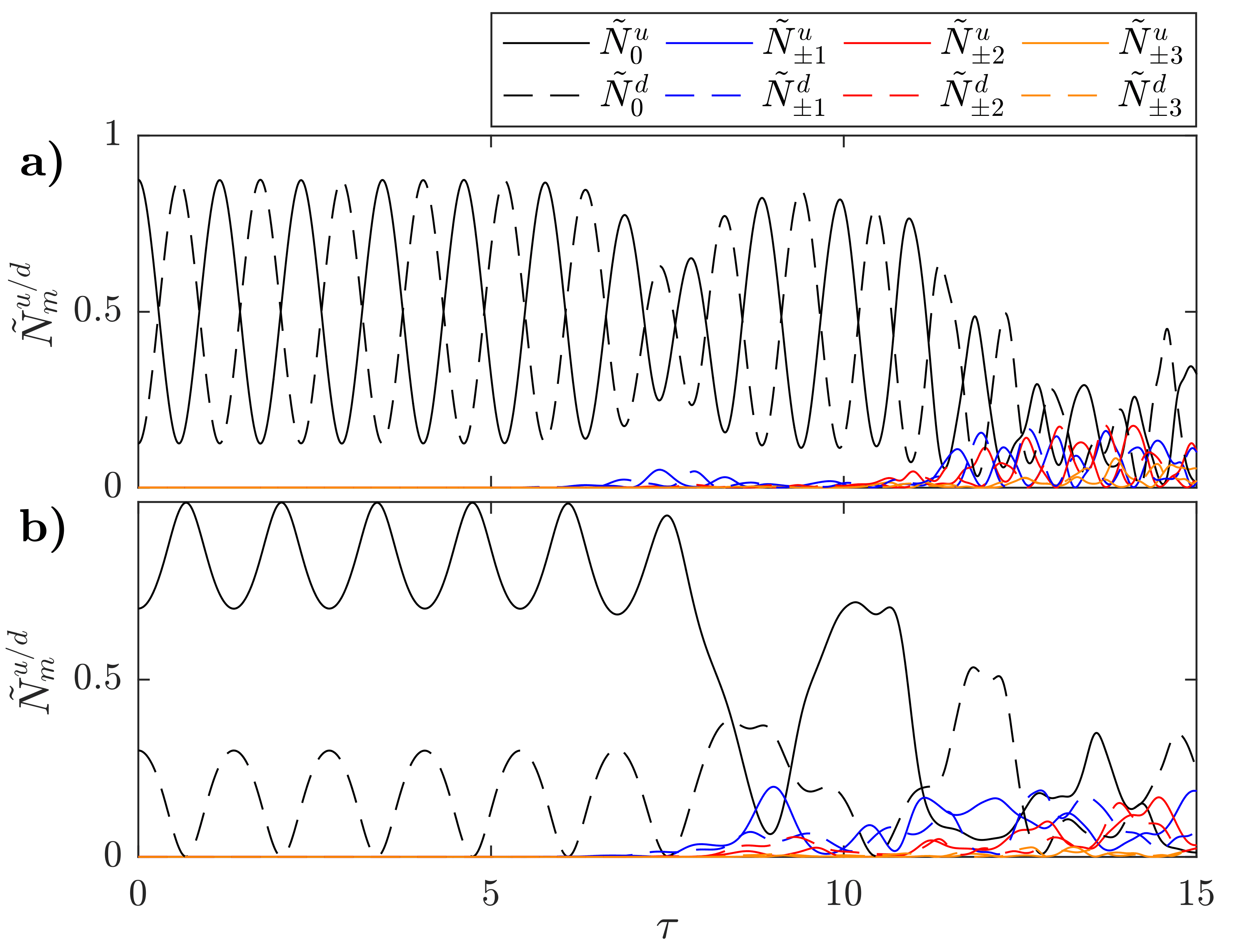}
	\caption{(Color online). Temporal evolution of the populations, $\tilde{N}_m^{u/d}=N_m^{u/d}/N$ with $N=4000$, for unstable Josephson oscillations a) $\kappa=4$, $\epsilon=3$, $z_0(0)=0.75$ (square in Fig.~\ref{FigMaps}(c)) and self-trapping b) $\kappa=1.5$, $\epsilon=3.5$, $z_0(0)=0.4$ (circle in Fig.~\ref{FigMaps}(b)).}\label{FigUnstable}
\end{figure}

The criteria for classification are the following. The stable regimes are those for which the population of the perturbed modes remains below $0.01$. For stable Josephson oscillations, the population imbalance of the main mode becomes zero at some point during time evolution, whereas in the stable self-trapping regime it does not. The decay time of the unstable regimes is defined as the time for which the total mode populations, $N_m$, of the main mode and the perturbation modes cross.

The Josephson oscillations and self-trapping dynamics decay into unstructured oscillations when higher order modes get excited. The system then remains in a state of non-periodic oscillations between the two rings that involves several modes. Fig.~\ref{FigUnstable} presents examples of these dynamics for unstable a) Josephson oscillations and b) self-trapping, corresponding to the square in Fig.~\ref{FigMaps}(c) and the circle in Fig.~\ref{FigMaps}(b), respectively. 

\begin{figure}[!tbp]
	\includegraphics[width=0.97\columnwidth]{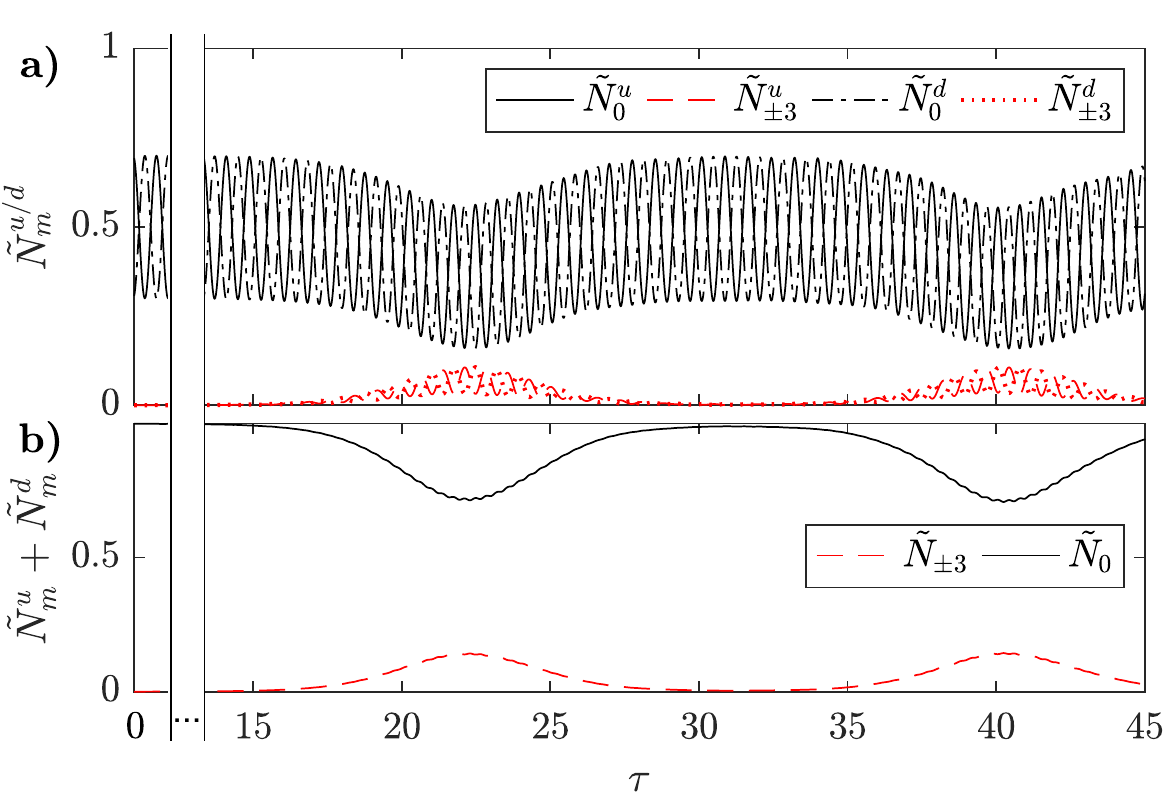}
	\caption{(Color online). Temporal evolution of semistable Josephson oscillations for $\kappa=4.5$, $\epsilon=1$, $N=4000$, and $z_0(0)=0.4$ (triangle in Fig.~\ref{FigMaps}(b)) for a) the populations in each mode and ring, $\tilde{N}_m^{u/d}=N_m^{u/d}/N$, and b) the total mode populations, $\tilde{N}_m^u+\tilde{N}_m^d$. Note that the time axis has a gap between $\tau=0$ and $\tau=15$ to show the relevant dynamics. }\label{FigJOsemistable}
\end{figure}

\begin{figure}[t]
	\includegraphics[width=1\columnwidth]{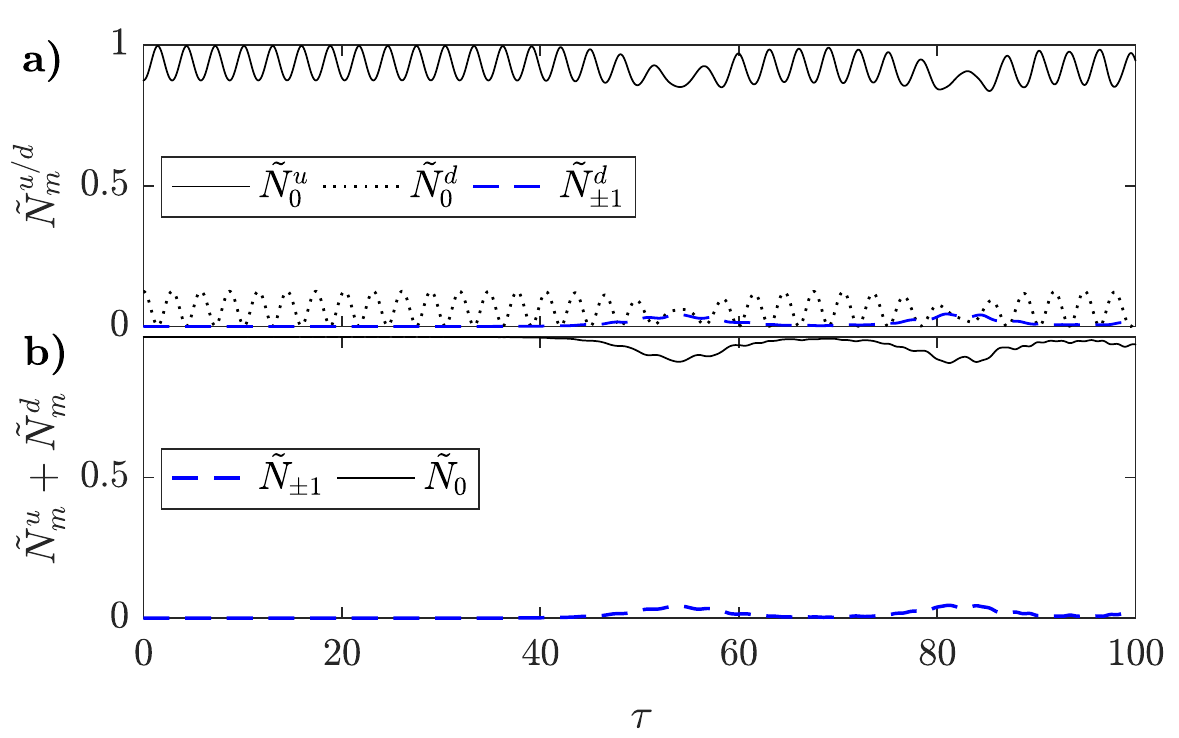}
	\caption{(Color online). Temporal evolution of semistable self-trapping for $\kappa=0.35$, $\epsilon=1.25$, $N=4000$ and $z_0(0)=0.75$ (rhombus in Fig.~\ref{FigMaps}(c)), for a) the populations in each mode and ring, $\tilde{N}_m^{u/d}=N_m^{u/d}/N$, and b) the total mode populations, $\tilde{N}_m^u+\tilde{N}_m^d$. The population of the mode $m=\pm1$ remains below $0.005$.}\label{FigSTsemistable}
\end{figure}

Close to the boundary between the stable and the unstable regimes, the system presents semistable Josephson oscillations and self-trapping. In these cases, the population of a single excited mode $\pm m$ grows and decays periodically, without destabilizing the dynamics of the main mode, $n=0$. The upper plots of Figs.~\ref{FigJOsemistable} and \ref{FigSTsemistable} show an example of semistable Josephson dynamics and semistable self-trapping dynamics, respectively. The lower plots of these figures show the corresponding total mode populations $\tilde{N}_m^u+\tilde{N}_m^d=(N_m^u+N_m^d)/N$, which present a pattern analogous to those shown by Bogoliubov excitations of the stationary state (see Fig.~\ref{comparison}). Therefore, the semistable dynamics can be understood as Bogoliubov excitations of the dynamical states modulated by tunneling.


\section{Conclusions}\label{SecConclusions}
In this work, we have investigated a Bose-Einstein condensate with repulsive interactions trapped in two rings in a stack configuration. The stability and dynamics of the BEC have been studied within mean-field theory and in terms of its OAM modes. For the case of a single mode equally populated in both rings and including small perturbations in other modes, we have derived a two-state model that predicts the regions of the parameter space supporting stable stationary states. This model also describes the dynamics of the system after destabilization, and characterizes accurately the features of the excitations. The analytical results of the two-state model have been contrasted with the numerical integration of the full model, finding a good qualitative and quantitative agreement.

Also, we have analyzed the dynamics of the system when a single OAM mode is populated with an arbitrary population imbalance between the two rings: the dynamical regimes of Josephson oscillations and self-trapping. The boundary condition between the two regimes has been analytically derived in terms of the population imbalance and the corresponding phase difference. We have found that the dynamics are equal for all OAM modes, and resemble the dynamics of a double-well system. By numerical analysis, we have characterized these dynamical regimes against perturbations in higher order OAM modes.



\vspace{-3.5mm}
\section{Acknowledgements}
We are grateful to Armengol Gasull for fruitful discussions. Also, EN, VA and JM acknowledge support through the Spanish Ministry of Science and Innovation (MINECO) (Contract No. FIS2017-86530-P)  and the Catalan Government (Contract No. SGR2017-1646). EN acknowledges financial support from MINECO through the grant PRE2018-085815. BJD acknowledges support through MINECO (Contract No. FIS2017-87534-P) and the Catalan Government (Contract No. 2017SGR-533).



%

\clearpage
\onecolumngrid
\section*{Appendix}
For the two-state model derived in Sec. \ref{SecTwoStateModel}, the behavior of the system around the critical points can be obtained by studying the eigenvalues of the Jacobian at the critical points. The Jacobian reads
\begin{equation}
\begin{aligned}
DF=
\begin{pmatrix}
\dfrac{\partial\dot{|\alpha_m|^2}}{\partial |\alpha_m|^2} & \dfrac{\partial\dot{\zeta}}{\partial |\alpha_m|^2}\vspace{1mm}\\
\dfrac{\partial\dot{|\alpha_m|^2}}{\partial \zeta} & \dfrac{\partial\dot{\zeta}}{\partial \zeta}
\end{pmatrix}=
\begin{pmatrix}
\gamma\left(8|\alpha_m|^2-N\right)\sin{2\zeta} & 3\gamma+4\gamma\cos{2\zeta}\vspace{2mm}\\
4\gamma |\alpha_m|^2\left(2|\alpha_m|^2-\frac{N}{2}\right)\cos{2\zeta} & -2\gamma\left(4\gamma |\alpha_m|^2-\frac{N}{2}\right)\sin{2\zeta}
\end{pmatrix}.
\end{aligned}
\end{equation}

For the critical point ($\zeta=a\pi$,$|\alpha_m|^2$),
\begin{equation}
DF(\zeta=a\pi,|\alpha_m|^2)=\begin{pmatrix} 0 & 7\gamma \\ 4\gamma |\alpha_m|^2 \left(2|\alpha_m|^2-\frac{N}{2}\right) & 0 \end{pmatrix},
\end{equation}
and the eigenvalues are
\begin{equation}
\lambda=\pm\sqrt{28\gamma^2|\alpha_m|^2\left(2|\alpha_m|^2-\dfrac{N}{2}\right)}.
\end{equation}
If condition (\ref{EqCondition1}) is fulfilled, and thus there are excitations, these eigenvalues are imaginary, and thus the stationary point is a center, the trajectories precede around it. 

For ($\zeta=(2a+1)\frac{\pi}{2}$,$|\alpha_m|^2$),
\begin{equation}
DF\left(\zeta=(2a+1)\dfrac{\pi}{2},|\alpha_m|^2\right)=\begin{pmatrix} 0 & -\gamma\\ -4\gamma |\alpha_m|^2\left(2|\alpha_m|^2-\frac{N}{2}\right) & 0 \end{pmatrix},
\end{equation}
with eigenvalues 
\begin{equation}
\lambda=\pm\sqrt{4\gamma^2|\alpha_m|^2\left(2|\alpha_m|^2-\dfrac{N}{2}\right)}.
\end{equation}
As before, if condition (\ref{EqConditionRestrictive}) is fulfilled, these stationary points are centers, and the trajectories precede around them. For ($\zeta$,$|\alpha_m|^2=0$),
\begin{equation}
DF(\zeta,|\alpha_m|^2=0)=\begin{pmatrix} -\gamma N\sin{2\zeta} & 3\gamma+4\gamma\cos{2\zeta} \\
0 & \gamma N\sin{2\zeta} \end{pmatrix}.
\end{equation}
As the eigenvalues are real and have opposite sign, $\lambda=\pm\gamma N \sin{2\zeta}$, the stationary point is a saddle point.

Similarly, for ($\zeta$,$|\alpha_m|^2=N/4$), 
\begin{equation}
DF\left(\zeta,|\alpha_m|^2=\dfrac{N}{4}\right)=\begin{pmatrix} \gamma N\sin{2\zeta} & 3\gamma+4\gamma\cos{2\zeta}\\ 0 & -\gamma N\sin{2\zeta} \end{pmatrix},
\end{equation}
with real eigenvalues of opposite sign $\lambda=\pm\gamma N\sin{2\zeta}$.

\end{document}